\documentclass[twocolumn,showpacs,preprintnumbers,amsmath,amssymb]{revtex4} 

\usepackage{graphicx}
\usepackage{dcolumn}
\usepackage{bm}
\usepackage{color} 

\usepackage[latin1]{inputenc}
\usepackage[english]{babel}
\usepackage{amssymb}
\usepackage{graphicx}

\newcommand{\be}{\begin{equation}}
\newcommand{\ee}{\end{equation}}
\newcommand{\mD}{\mathcal{D}}
\newcommand{\mU}{\mathcal{U}}
\newcommand{\mL}{\mathcal{L}}

\newcommand{\mE}{\mathcal{E}}
\newcommand{\la}{\langle}
\newcommand{\ra}{\rangle}

\newcommand{\sign}{{\rm sign}}
\newcommand{\Erfc}{{\rm Erfc}}
\newcommand{\bx}{{\bf x}}
\newcommand{\bu}{{\bf u}}

\begin{document}

\title{Lagrangian formulation of turbulent premixed combustion}

\author{Gianni Pagnini}
\author{Ernesto Bonomi}
\affiliation{
CRS4, Polaris Building 1, 09010 Pula, Italy}

\date{\today}

\begin{abstract}
The Lagrangian point of view is adopted to study turbulent premixed combustion. 
The evolution of the volume fraction of combustion products is established by the Reynolds transport theorem. 
It emerges that the burned-mass fraction is led by the turbulent particle motion, by the flame front velocity, 
and by the mean curvature of the flame front. 
A physical requirement connecting particle turbulent dispersion and flame front velocity is obtained 
from equating the expansion rates of the flame front progression and of the unburned particles spread. 
The resulting description compares favorably with experimental data. 
In the case of a zero-curvature flame, with a non-Markovian parabolic model for turbulent dispersion, 
the formulation yields the Zimont equation extended to all elapsed times and fully determined by turbulence characteristics. 
The exact solution of the extended Zimont equation is calculated and analyzed to bring out different regimes.
\end{abstract}

\maketitle
Turbulent premixed combustion is a challenging scientific field involving nonequilibrium
phenomena and playing the main role in important industrial issues such as energy production
and engine design. 

A Lagrangian point of view is here adopted, leading to a description of turbulent premixed
combustion which takes into account, for all elapsed times, the turbulent dispersion, the
volume consumption rate of reactants, and the flame mean curvature.
The proposed approach generalizes and unifies classical literature approaches
that are based on
the so-called level-set method \cite{sethian_etal-arfm-2003}
or are based on the Zimont balance equation,
originally hinted at by Prudnikov \cite{prudnikov-1964},
also known as Turbulent Flame Closure model  \cite{zimont-etfs-2000}.
Moreover, the proposed formulation has the striking property to be compatible with every 
type of geometry and flow
in an easier and more versatile way than previous approaches,
and it emerges to be easily modifiable to include more detailed and correct physics.
It is worth recalling that the
Zimont equation was introduced on the basis of experimental observations and that a great
effort has been undertaken to give a deeper theoretical foundation to it
\cite{zimont-etfs-2000,lipatnikov_etal-pecs-2002,lipatnikov-2004,lipatnikov_etal-pf-2005}.
The present Lagrangian formulation constitutes a reliable theoretical support for the Zimont
combustion model.

The process of turbulent premixed combustion is mainly characterized by flame propagation 
towards the unburned region and turbulent dispersion of the resultant product
particles. The combustion process is described by a single dimensionless scalar observable,
denoted as \emph{average} progress variable, $0\leq c(\bx,t)\leq 1$, 
and representing the burned-mass fraction, i.e., the fraction of burned particles which are located in $\bx$ at time $t$. 
The value $c(\bx,t)=1$ describes the presence of only products
and the value $c(\bx,t)=0$ describes the presence of only reactants.
To avoid unnecessary mathematical difficulties, we consider a constant-density mixture and a
zero-mean turbulent velocity field. Molecular diffusion is also neglected.

In this Letter, the fresh mixture is intended to be a population of particles in turbulent
motion that, in a \emph{statistical} sense, change from reactant to product when their average positions are hit by the flame. 
Let $\Omega(t)$ be the portion of space surrounded by the flame surface; then
those particles with average position $\la \bx \ra \in \Omega(t)$ are marked as burned particles.
The occurrence in $\bx$ at time $t$ of a particle transit is described by a probability 
density function (PDF). Let $p(\bx;t|\bx_0)$ be the PDF associated with a particle
displacement 
where $\bx_0$ is the initial condition of a Lagrangian trajectory
and, without loss of generality, let $t=0$ be the ignition instant. 
With the assumption that particle trajectories are not affected by the chemical
transformation, the average progress variable $c(\bx,t)$ turns out to be defined as the 
superposition 
of PDFs of burned particles, i.e., those $p(\bx;t|\bx_0)$ with $\la \bx \ra \in \Omega(t)$.
For a zero-average velocity field, the particle average position is $\la \bx \ra =\bx_0$
and then
\be
c(\bx,t)=\int_{\Omega(t)} p(\bx;t|\bx_0) \, d\bx_0  \,.
\label{def:progressvariable}
\ee
The evolution law for the progress variable $c(\bx,t)$ is obtained applying 
Reynolds transport theorem to (\ref{def:progressvariable}) which gives
\be
\frac{\partial c}{\partial t} =
\int_{\Omega(t)}\frac{\partial p}{\partial t} \, d\bx_0 
+
\int_{\Omega(t)}\nabla_{\bx_0} \cdot [\bu(\bx_0,t) p(\bx;t|\bx_0)] \, d\bx_0 \,,
\label{rtt}
\ee
where
$\nabla_{\bx_0}$ is the gradient with respect to $\bx_0$ and 
$\bu(\bx,t)$ is the expansion velocity field of $\Omega(t)$.

Let the turbulent dispersion be represented by the general evolution equation
\be
\frac{\partial p}{\partial t}=\mE_{\bx} [ \, p \, ] \,,
\quad p(\bx;0|\bx_0)=\delta(\bx-\bx_0) \,,
\label{evolp}
\ee
where the spatial operator $\mE_{\bx}[\cdot]$ includes the particle 
displacement statistics such as the variance {\small $\sigma^2(t)=\la||\bx-\bx_0||^2\ra/3$}.

Equation (\ref{rtt}) is also governed by the volumetric expansion of $\Omega(t)$.
This expansion is connected with the \emph{consumption rate} 
that in a general form is set to be
\be
\bu(\bx,t)=\mU(\kappa,t) \, \hat{n} \,,
\quad \hat{n}=-\frac{\nabla c}{||\nabla c||} \,,
\label{mbU}
\ee
where $\kappa(\bx,t)=\nabla \cdot \hat{n}/2$ denotes the local mean \emph{curvature}. 
Since molecular processes are neglected, the initial burning speed must be zero, i.e. $\mU(\kappa,0)= 0$.
From (\ref{mbU}) the location of the flame surface follows to be
\be
\mL_f(t)=\mL_0 + \int_0^t \bu(\mL_f,\tau) \, d\tau \,.
\label{Lf}
\ee
Finally, inserting (\ref{evolp}) and (\ref{mbU}) in (\ref{rtt}) gives
\begin{eqnarray}
\frac{\partial c}{\partial t} = \mE_{\bx}[\, c \,] + \int_{\Omega(t)} \bu \cdot \nabla_{\bx_0} p \ d\bx_0 \hspace{3truecm}\nonumber \\
+ \int_{\Omega(t)} p \, \left\{
\frac{\partial \mU}{\partial \kappa}\, \nabla_{\bx_0}\kappa \cdot \hat{n}+
2 \mU(\kappa,t) \kappa(\bx_0,t) \right\} d\bx_0 \,.
\label{reynolds}
\end{eqnarray}
The evolution of the progress variable is then led by three factors:
turbulent motion, displacement speed of the contours of $c(\bx,t)$ and their mean curvature.
It is worth remarking that (\ref{reynolds}) cannot be reduced to the most widely used 
front propagation equations \cite{xin-sr-2000},  
and, since (\ref{reynolds}) follows from the exact Lagrangian definition (\ref{def:progressvariable}),
none of them is physically correct to model turbulent premixed combustion.

When particle motion is neglected, products and reactants turns out to be frozen, 
$\partial p/\partial t=0$, so that $p \to \delta(\bx-\bx_0)$ and
$\int_{\Omega(t)} \nabla_{\bx_0}\cdot[\bu(\bx_0,t)p] d\bx_0=-\bu \cdot \nabla c$. Here 
the identity $\nabla p=-\nabla_{\bx_0}p$ has been used. In this limit case, using (\ref{mbU}), 
Eq. (\ref{rtt}) reduces to
\be
\frac{\partial c}{\partial t}= \mU(\kappa,t) \, ||\nabla c|| \, ,\hspace{0.3cm}\ c(\bx,t)=
\left\{
\begin{array}{cc}
1 \,, & \mbox{if } \bx \in \Omega(t) \,,     \\
\\
0 \,, & \mbox{otherwise} \,,
\end{array}
\right.
\label{sethian}
\ee
the celebrated Hamilton--Jacobi equation stated by Sethian \cite{sethian_etal-arfm-2003} 
to track the flame front surrounding the burned volume $\Omega(t)$,
which is related to the G-equation \cite{peters-jfm-1999} and 
to the Kardar-Parisi-Zhang equation \cite{kerstein_etal-pra-1988}.
Equation (\ref{sethian}) can be now interpreted as a consequence of Reynolds transport theorem.

When the normal $\hat{n}$ to the contours of the progress variable is assumed constant, 
then the mean curvature $\kappa$ is zero. 
Assuming an homogeneous, isotropic and stationary turbulence, for the Lagrangian PDF it holds
$\nabla p=-\nabla_{\bx_0}p$ and setting $\mU(0,t)=\mU(t)$ 
formula (\ref{reynolds}) turns out to be
\be
\frac{\partial c}{\partial t}=\mE_{\bx}[\,c\,]   
+\mU(t) \, || \nabla c || \,.
\label{zimont}
\ee

\begin{figure}[htp]
\begin{center}
\includegraphics[width=8cm,height=5cm,angle=0]{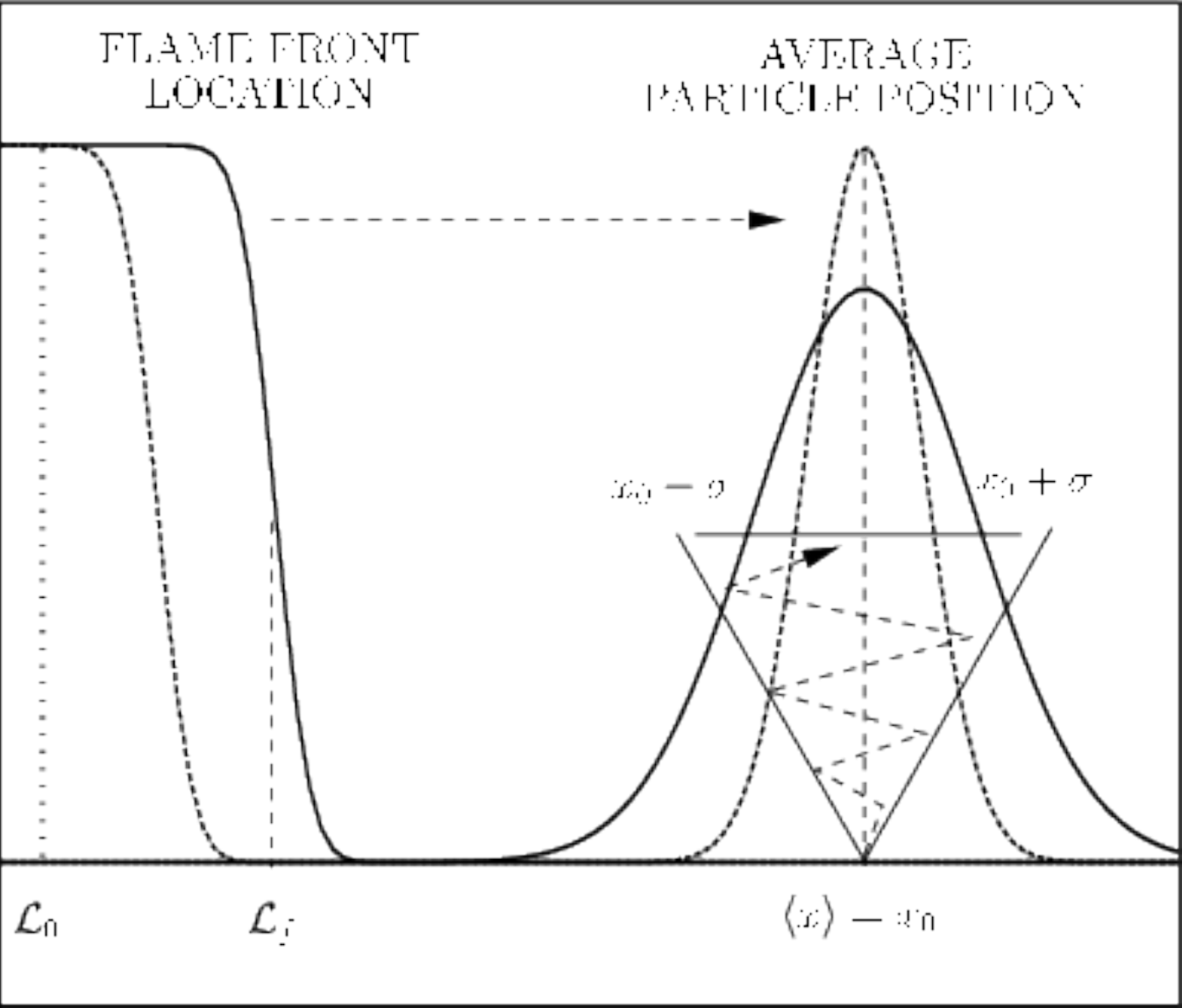}
\caption{Illustration of the forward motion of the flame front from the initial position $\mL_0$ to $\mL_f(t)$ 
and the expansion of the particle quadratic mean $\sigma$ after a forward plus backward random walk 
around the average position $\la x \ra=x_0$.}
\vspace{-1truecm}
\label{picture}
\end{center}
\end{figure}
It must be observed that in (\ref{zimont}) the turbulent dispersion and the flame expansion enter 
in the progress variable evolution with the particle displacement 
variance $\sigma^2(t)$ and $\mU(t)$, that are so far independent. 
However, in a proper combustion model they have to be mutually related since,
when molecular processes are neglected,
the flame front has to be solely fueled and carried by the turbulent dynamics of the reacting environment.   
To formulate a correspondence between particle spread and flame progression, 
the expansion rate of the quadratic mean of particle displacement $r(\tau,t)=\sigma(t+\tau)/\sigma(t)$ 
is taken equal to the expansion rate of the flame front progression $\Delta_f(t)=\mL_f(t)-\mL_0$.
In the mechanism here proposed, while the combustion evolves moving along the outward flame front normal, 
see Fig. \ref{picture},
the expansion of $\sigma$ is statistically related to the random oscillation of a particle moving 
forward and backward around its mean position.
As a consequence, a \emph{half} time step $\tau$ is necessery to the forward-moving flame front 
to have the \emph{same} expansion rate than an unburned particle oscillating around its mean position: 
$r(\tau,t)=\Delta_f(t+\tau/2)/\Delta_f(t)$. 
Finally, noting that $r(0,t)=1$ and performing the limit $\tau \to 0$, the joined process satisfies
\be
\frac{1}{\mL_f(t)-\mL_0}\frac{d \mL_f}{d t}=\frac{2}{\sigma(t)}\frac{d \sigma}{d t}=
2 \left.\frac{\partial r}{\partial \tau}\right|_{\tau=0} \,.
\label{ratio0}
\ee
Let us introduce the function
\be
\mD(t)=\frac{1}{2}\frac{d \sigma^2}{d t}=\int_0^t B_L(\tau) \, d\tau \,,
\label{D}
\ee
where $B_L(t)$ is the Lagrangian velocity autocorrelation function.
It is worth remarking that definition (\ref{D}) is in agreement with the exact Taylor formula
$\sigma^2(t)=2\int_0^t (t-\tau)B_L(\tau) \, d\tau$
that includes all turbulent dispersion regimes from the ballistic to the diffusive one
passing through the inertial range.
Using definitions (\ref{Lf}) and (\ref{D}), identity (\ref{ratio0}) can be written in terms of $\mU$ and $\mD$ 
and yields
\be
\frac{\int_0^t \mU(\xi) \, d\xi}{\mU(t)} \,
\frac{\mD(t)}{\int_0^t \mD(\xi) \, d\xi}=1 \,.
\label{ratio}
\ee
It follows that
$\int_0^t [\mD(t) \mU(\xi) - \mU(t) \mD(\xi)] \, d\xi =0$ for any $t\geq 0$.
From the monotonicity of both $\mU$ and $\mD$, since 
they are non-negative functions and $\mU(0)=\mD(0)=0$, the following equality holds
\be
\frac{\mD(\xi)}{\mU(\xi)}=\frac{\mD(t)}{\mU(t)} \,,
\quad 0 \le \xi \le t \, ,
\label{rapporto1}
\ee
so that, within the integration interval of $\xi$, the ratio $\mD(\xi)/\mU(\xi)$ must be constant.
Moreover, for $t \to \infty$ both $\mD(t)$ and $\mU(t)$ are bounded, 
$\mD(\infty)=\mD_{eq}$ and $\mU(\infty)=\mU_{eq}$;
therefore, such a constant is equal to
\be
\frac{\mD(t)}{\mU(t)}=\frac{\mD_{eq}}{\mU_{eq}}=\lambda \,,
\quad t \ge 0 \,.
\label{lambda}
\ee
Identity (\ref{lambda}) states that, by using definition (\ref{D}), 
the whole evolution of the combustion process is \emph{solely} established by $\sigma^2(t)$,
and it constitutes a new result in literature \cite{xin-sr-2000}. Moreover, it determines not only the relation between the combustion drift
and the background turbulent dispersion, 
but also the temporal evolution of the flame front location $\mL_f(t)$ as it follows 
\be
\mL_f(t)=\mL_0 + \frac{\sigma^2(t)}{2 \lambda} \,.
\label{LR}
\ee
A similar result
was already sketched by Biagioli \cite{biagioli-ctm-2006} in the study of strongly swirled flows, 
but valid only for asymptotic long times. 

\begin{figure}[htp]
\begin{center}
\includegraphics[width=8cm,height=6cm,angle=0]{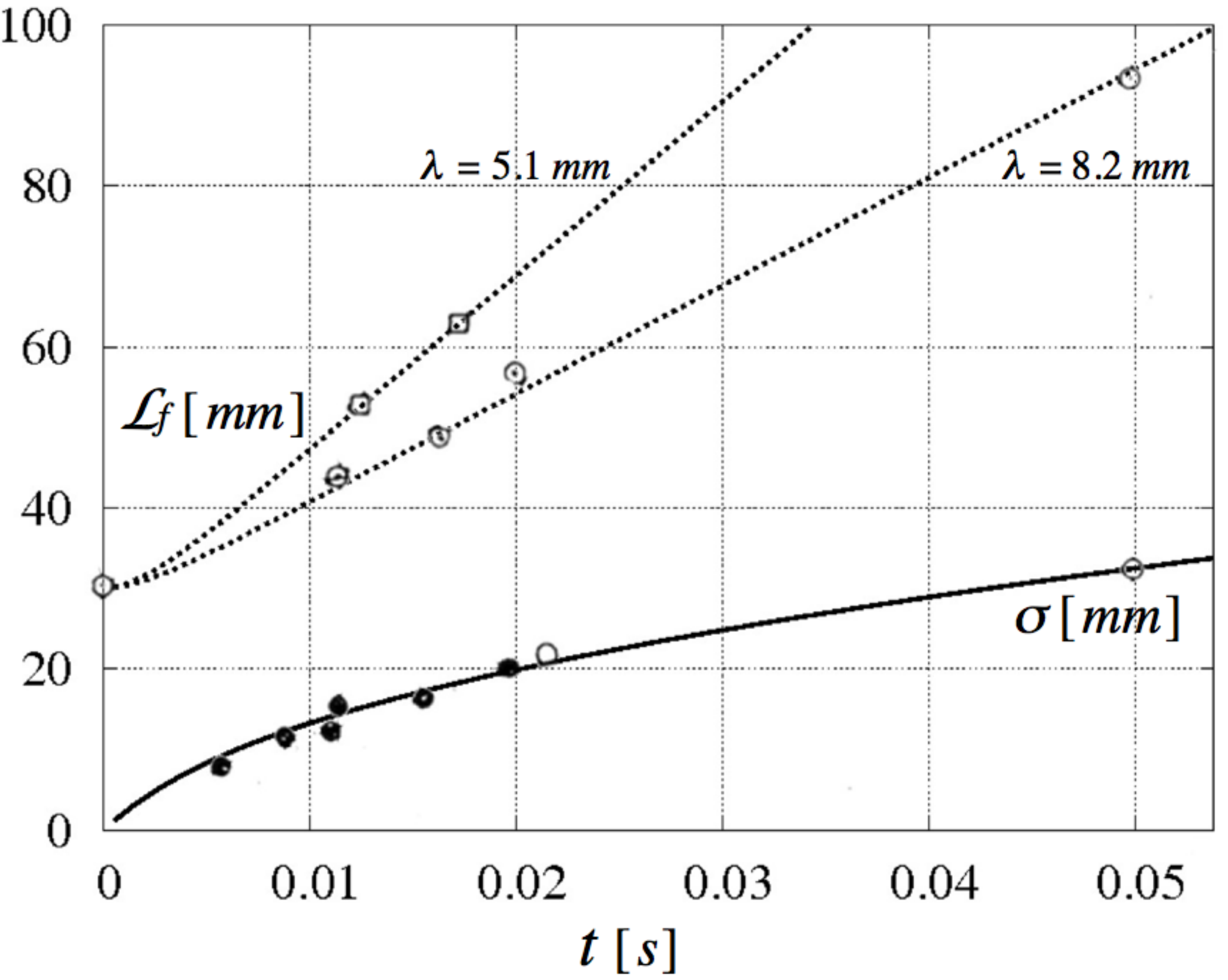}
\caption{
Experimental validation of the propagation law (\ref{LR}). Lines are the plots
of the analytic $\sigma$ and the predicted flame front positions $\mL_f$.
The values $\lambda=5.1\,  \rm mm$ and $\lambda=8.2\, \rm  mm$ 
correspond to two different equivalence ratios $F=0.68$ and $F=0.56$, respectively.  
}
\vspace{-1truecm}
\label{validazione}
\end{center}
\end{figure}

To validate the goodness of the physical argument that brought to expression (\ref{LR}), 
an experimental result discussed in the literature is used where $\sigma$ and $\mL_f$ 
were simultaneously measured.
The data come from figure $21$ in Ref. \cite{lipatnikov_etal-pecs-2002}
and they are attributed to unavailable measurements \cite{sherbina-mipt-1958}.
In the experimental setup, the gas mixture flows at the steady velocity $V=26.0$ $\rm ms^{-1}$ 
and the data acquisition was performed at fixed distances $x[\rm mm]$ from an origin.
The measurement locations $x[\rm mm]$ have been converted in elapsed times $t[\rm s]$ by $t=x/V$ and 
$\mL_f$ and $\sigma$ have been considered as measured in a reference frame 
in translation with the flow.
Figure \ref{validazione} shows the fit between the measurements of $\sigma$ and Taylor formula,
where the exponential autocorrelation function $B_L(t)=\la u'^2 \ra \exp(-t/T_L)$ is assumed 
with $\mD_{eq}=\la u'^2 \ra\, T_L=11\times10^3 \rm mm^2\, {\rm s^{-1}}$ and 
$T_L=2\times10^{-3}\, \rm s$. 
Here $T_L=\la u'^2 \ra^{-1}\int_0^t B_L(\tau)\, d\tau$ denotes the Lagrangian integral timescale.
Substituting the resulting analytic $\sigma^2$ in Eq. (\ref{LR}), 
the flame front position $\mL_f$ is correctly predicted. 
The experiment was performed twice with the same turbulence characteristics
but with two different equivalence ratios $F$
(that is the ratio of the fuel-to-oxidizer ratio to the stoichiometric fuel-to-oxidizer ratio):
$F=0.68$ and $F=0.56$.  

Let us consider now the simple non-Markovian parabolic model: $\mE_{\bx}[\, p\,] \equiv \mD(t)\nabla^2 p$. 
Equation (\ref{zimont}) becomes the extension to all elapsed times
of the familiar Zimont equation \cite{zimont-etfs-2000},
which was historically formulated in the asymptotic regime $t \gg T_L$ with $\mD(t)$ tending to $\mD_{eq}$ and $\mU(t)$ to $\mU_{eq}$.
A critical review about it can be found in \cite{lipatnikov_etal-pecs-2002,lipatnikov_etal-pf-2005}.

Despite the restriction $\kappa=0$, 
other extensions of (\ref{zimont}) to the initial regime $t < T_L$ are currently applied 
in engineering applications to
study transient and geometrical effects in the developing phase
of the flame \cite{lipatnikov_etal-cst-2000} with practical
fallouts in the design of spark-ignition engines \cite{lipatnikov_etal-je-1997}.

With $\kappa=0$, the study of Zimont equation is reduced
to a one-dimensional problem along the normal direction to the flame front. 
The front speed (\ref{mbU}) becomes $u(x,t)=- \, \mU(t) \, \sign(\partial c/\partial x)$.
Let the portion of space surrounded by the flame be  
$\Omega(t)=[\mL_L(t),\mL_R(t)]$, where $\mL_L$ and $\mL_R$ are the flame front positions defined in (\ref{Lf}) on the left and on the right of the ignition point, respectively.
Then the exact solution is
\be
c(x,t)=
\frac{1}{2}
\left\{\Erfc\left[\frac{x-\mL_R(t)}{\sqrt{2} \, \sigma(t)}\right]
- \Erfc\left[\frac{x-\mL_L(t)}{\sqrt{2} \, \sigma(t)}\right]
\right\} \,. 
\label{c}
\ee
By setting $x=z-z_0$ and $\mL_R(t)=z_f(t)-z_0$, in the limit 
$z_0 \to -\infty$ the progress variable becomes $c(z,t)=\Erfc\left[(z-z_f(t))/(\sqrt{2} \, \sigma(t)) \right]/2$, 
that is in agreement with several experimental results \cite{lipatnikov_etal-pecs-2002,lipatnikov_etal-pf-2005}.
In particular, formula (\ref{LR}) can be plugged into solution (\ref{c}) of the \emph{extended} Zimont equation.
With this model, now fully determined for all elapsed times by $\sigma^2$, 
it is possible to perform a generally valid analysis
on the flame \emph{enhancement} and \emph{quenching}
and bring out the different regimes of the combustion process.
\begin{figure}[htp]
\begin{center}
\includegraphics[width=8cm,height=6cm,angle=0]{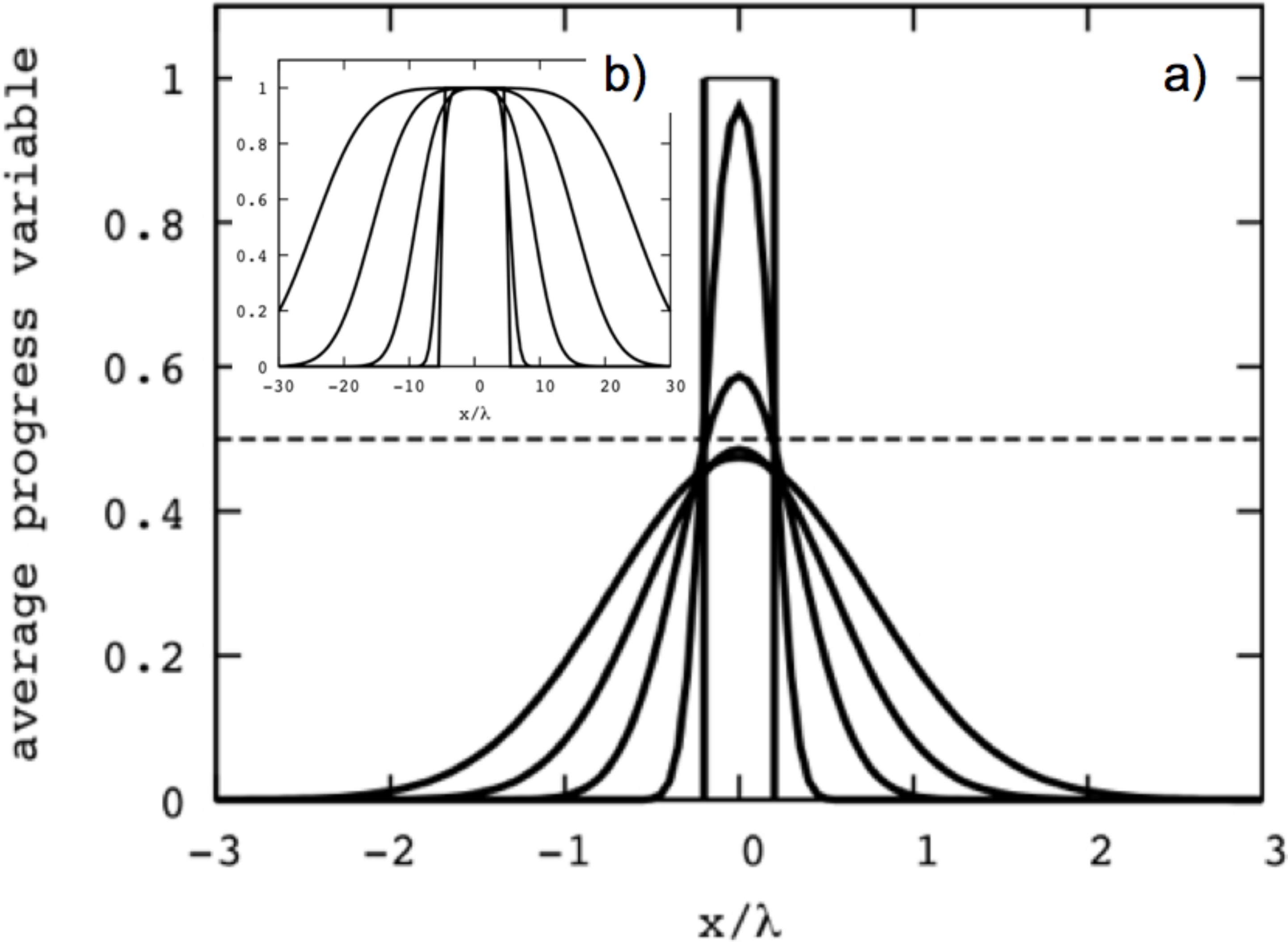}
\caption{Evolution of $c(x,t)$ starting from a fully burned zone: 
a) $\mL_0= \lambda/5$ and $t=0$, $0.01 T_L$, $0.03 T_L$, $0.05 T_L$, $0.07 T_L$;
b) $\mL_0= 5 \, \lambda$ and $t=0.01 T_L$, $0.1 T_L$, $0.3 T_L$, $0.5 T_L$, $0.7 T_L$. 
In a), the dashed line indicates the quenching threshold $c=0.5$.} 
\vspace{-1truecm}
\label{plot}
\end{center}
\end{figure}
The two examples of Fig. \ref{plot} display the evolution of $c(x,t)$ with the initial fully 
burned zone bounded by different flame front conditions: $\mL_0= \lambda/5$ 
and $\mL_0=  5 \, \lambda$.
For the plots, we have 
set $\la u'^2 \ra=1$, $T_L=1$, $\lambda=0.1$.
Observe in Fig. \ref{plot}a that, after the initial ignition instant, 
turbulent diffusion mixes together products and reactants so that 
the progress variable takes a bell-shape profile and
in the central zone $c(x,t)$ is drained by the dispersion of
the resulting product particles. 
If $c(\bx,t)$ becomes less than a threshold value, 
quenching can be assumed.
The larger is the ignition region the weaker is the draining, and,  as illustrated in Fig. \ref{plot}b,
no-draining occurs for an initial value $\mL_0$ larger than a critical lenghtscale $\mL_{0c}$.
The lenghtscale $\mL_{0c}$ can be determined by the budget between the spread of the particle distribution, 
driven by $\mD(t)$, and the expansion of $\Omega(t)$, driven by $\mU(t)$. 
Introducing a pseudo diffusion coefficient for the combustion $\mU(t) \mL_0$,
the budget emerges to be given by the ratio $\mU(t)\mL_0/\mD(t)=\mL_0/\lambda$.
It emerges from this analysis that $\mL_{0c}=\lambda=\mD_{eq}/\mU_{eq}=\sqrt{\mD_{eq} \mathcal{T}}$, 
where $\mathcal{T}=\mD_{eq}/\mU^2_{eq}$ is a determination of the characteristic reaction time. 
When $\mL_0 \ll \lambda$, the flame rapidly quenches after the initial ignition, see Fig. \ref{plot}a;
when $\mL_0 \gg \lambda$, the flame is immediately able to sustain itself, see Fig. \ref{plot}b, and 
the combustion propagates. 
This dependence on the size of $\mL_0$ is the same as the one theoretically estabilished for the 
Kolmogorov--Petrovskii--Piskunov models \cite{zlatos-jams-2006}. 
In the intermediate case  $\mL_0 \simeq \lambda$, when
turbulence is not strong enough to drain $c(x,t)$ below the quenching threshold,
$c(x,t)$, after an initial fall, is refilled and resustained by the combustion.
This last behavior identifies the regime in which the process is dependent on the initial condition. 
Actually, considering the right semiaxis on the flame front location $x=\mL_R(t)$, 
the two contributions of the $\Erfc$-functions in (\ref{c}) 
become $1$ and $\Erfc\{\sqrt{2} \,[\mL_0/\sigma + \sigma/(2 \lambda)]\}$, respectively.
Since $\sigma(t)$ is monotonic and increasing with $\sigma(0)=0$,
there exist two timescales $\tau_0$ and $\tau_*$ defined by
$\sigma(\tau_0)=\mL_0$ and $\sigma(\tau_*)=2 \lambda$, respectively,
which can be estimated from the scaling laws of $\sigma^2(t)$.
For short times, $t \ll T_L$, when $\sigma^2 \simeq \la u'^2 \ra t^2$,
it turns out that $\tau_0=\mL_0/\la u'^2 \ra^{1/2}$.
After this transient regime which is dependent on the initial condition, 
the process tends asymptotically to be self-similar.
The timescale $\tau_*$ turns out to be $\tau_*=2 (\la u'^2 \ra/\mU_{eq}^2)^{1/2} T_L$, 
if $\tau_* \ll T_L$, or $\tau_*=2 \lambda/\mU_{eq}$, if $\tau_* \gg T_L$.
This means that when $t < \tau_0 < \tau_*$ then
$\mL_0/\sigma > \sigma/(2\lambda)$ while when
$t > \tau_*$ then $\mL_0/\sigma < \sigma/(2\lambda)$.
So when $t < \tau_*$ the second $\Erfc$-function is strongly variable, while when
$t > \tau_*$ its argument begins to grow 
and the $\Erfc$-function tends rapidly to zero. 
Finally, when $t < \tau_*$ the average progress variable profile is
not self-similar, because each $\Erfc$-function in (\ref{c}) has its own self-similarity variable,
while for $t > \tau_*$ it is asymptotically self-similar with
respect to the self-similarity variable $(x-\mL_R(t))/\sigma(t)$.
Then the timescale $\tau_*$
separates the initial not self-similar transient regime from the 
asymptotically self-similar regime.
For long elapsed times, in the reference frame moving with the flame front, 
the smoothing effect of the turbulence spread decreases and asymptotically vanishes, making
the profile of $c(x,t)$ steeper. 
The same argouments above can be applied to the left semiaxis.
In addition, it is possible to expand the present formulation including the advecting mean velocity 
by following the same arguments formulated in \cite{berti_etal-epl-2008} where
$\mD_{eq}$ is changed in an effective diffusion coefficient with a nontrivial dependence on the mean velocity field. 
Then, by replacing (\ref{evolp}) and (\ref{D}) with a Lagrangian dispersion model for shear or cellular flow, 
the dependence of $\mL_{0c}$ 
on the mean velocity may be estimated as in \cite{constantin_etal-cpam-2001,vladimirova_etal-ctm-2003}.

GP is supported by the Sardinian Regional Authority 
(PO Sardegna FSE 2007-2013, L.R. 7/2007). Professor A.N. Lipatnikov is acknowledged for providing Ref. \cite{prudnikov-1964}
and helping to discover a misprint in Fig. 21 \cite{lipatnikov_etal-pecs-2002} and 5.32 \cite{prudnikov-1964}.

\bibliography{abbr,combustion-biblio}
\end{document}